\documentclass[prl,aps,twocolumn,showpacs]{revtex4}
\usepackage{epsfig}
\usepackage{graphicx}
\usepackage{textcomp}

\begin{document}
\title{Noise enhanced persistence in a biochemical regulatory network with feedback control}
\author{Michael Assaf and Baruch Meerson}
\affiliation{Racah Institute of Physics, Hebrew University of
Jerusalem, Jerusalem 91904, Israel} \pacs{87.16.Yc, 87.15.Ya,
87.16.Xa, 05.40.-a}
\begin{abstract}

We find that discrete noise of inhibiting (signal) molecules can
greatly delay the extinction of plasmids in a plasmid replication system:
a prototypical biochemical regulatory network.
We calculate the probability distribution of the metastable state of
the plasmids and show on this example that the reaction rate
equations may fail in predicting the average number of regulated
molecules even when this number is large, and the time is much shorter than
the mean extinction time.
\end{abstract}
\maketitle

Many molecular species that control genetic regulatory networks are
present in low concentrations. The resulting fluctuations in
reaction rates may cause large random variations in the
instantaneous intracellular concentrations of molecular species
which, in their turn, may have important consequences in biological
functioning. This and related topics have attracted much recent
interest from the biology and physics communities
\cite{berg,guptasarma,arkin,paulsson,thattai,kaern,samoilov,friedman}.
Intracellular processes are often regulated via negative feedback by
signal molecules. It was assumed in the past that noise in the
signal component would randomize control of the regulated component.
More recently, it has been shown that this noise may actually
\textit{enhance} the robustness of the regulated component, bringing
the variation of its probability distribution below the Poissonian
limit \cite{paulsson}. Here we report a previously unexplored
dramatic impact the noise can have on the \textit{persistence} of
the regulated component in systems with negative feedback control.
Following Paulsson \textit{et al.} \cite{paulsson}, we will consider
a minimal two-component copy number control (CNC) model that, on the
one hand, includes standard intracellular processes and, on the
other hand, provides an adequate description to CNC of bacterial
plasmids. Plasmids are extra-chromosomal DNA molecules (typically,
circular and double-stranded) that are capable of autonomous
replication. They undergo intracellular dynamics of the birth-death
type with decay (mostly dilution by cell division) and autocatalytic
production inhibited by signal molecules. If there is no penetration
of new plasmids into the cell, a rare sequence of multiple decay
events will ultimately drive the plasmid population to extinction.
It may even cause the death of the cell if the plasmid contains a
vital gene. We will combine analytical and numerical approaches to
show that noise in the number of signal molecules can greatly delay
the plasmid extinction. We will also calculate analytically the probability distribution function (PDF) of the metastable state of the regulated molecules and show that widely used deterministic reaction rate
equations (RRE) may fail in predicting the average number of plasmids even
when this number is large. These remarkable effects do not require
an unusual molecular distribution and occur due to rare events when the
number of signal molecules is very small.

\textit{Model.} Consider a double negative-positive
feedback loop with plasmids denoted by $X$ and signal molecules
denoted by $S$. The plasmids promote the production of the signal
molecules, whereas the signal molecules inhibit the autocatalytic production of the
plasmids. The
RRE for the average concentrations of the two species are \cite{paulsson}:
\begin{eqnarray}
\left\{\begin{array}{rcl}\dot{X}&=&X \Psi(S/A)-X\,,\vspace{2mm}\\
\dot{S}&=&\alpha X - \beta S\,,
\end{array}\right.\label{meanfield}
\end{eqnarray}
where $\Psi(S/A)$ is a nonlinear and monotone decreasing
function of $S$, $\Psi(0)>1$, the parameter $A$ specifies
the inhibition strength of the signal molecules $S$, and time
and the rates are rescaled by the decay rate of the plasmids.
Equations~(\ref{meanfield}) have an
attracting fixed point $(\bar{X},\bar{S})$
[where $\bar{X}=\bar{S}/\eta$, $\Psi(\bar{S}/A)=1$, and $\eta=\alpha/\beta$] and an unstable fixed point
$(X_0,S_0)=(0,0)$. According to the RRE, the system would stay in
the $(\bar{X},\bar{S})$ state forever. The underlying \textit{stochastic} process,
however, behaves quite differently. A large enough fluctuation
ultimately depletes the plasmid population. The state with no plasmids is an absorbing state,
as the probability of escape from it is zero. Therefore, the $(X_0,S_0)$ state is
actually stable, whereas the stable fixed point $(\bar{X},\bar{S})$
of the deterministic model is \textit{metastable}. The mean extinction time (MET): the mean time it takes this stochastic process to reach
the absorbing state  is expected to be exponentially long in the (presumably large) average number of plasmids
in the metastable state, see \textit{e.g.}, Refs. \cite{vankampen,gardiner,AM}.

To account for the stochastic effects, consider a chemical master
equation (CME) that describes the evolution of the
probability $P_{m,n}(t)$ of having, at time $t$, $m$ plasmids and $n$
$S$-molecules. For $m,n\geq 1$ the CME is
\cite{paulsson}
\begin{eqnarray}
\hspace{-7mm}\dot{P}_{m,n}&=&(E_m^{-1}-1)g_{m,n}P_{m,n}+(E_m^1-1)m
P_{m,n}\nonumber\\
&+&\alpha m (E_n^{-1}-1)P_{m,n}+\beta(E_n^{1}-1)n
P_{m,n}\,,\label{fullmaster}
\end{eqnarray}
where $E_n^{j}f(n)=f(n+j)$ and
$g_{m,n}=m \Psi(n/A)$ \cite{effective}. Let us denote by $P_{n|m}$ the probability of
having, at time $t$, $n$ $S$-molecules conditioned on having $m$
plasmids, and by $\pi_m$ the probability of having $m$ plasmids
regardless of the number of $S$-molecules. We substitute the
identity $P_{m,n}=P_{n|m}\pi_{m}$ into Eq.~(\ref{fullmaster}) and
sum over all $n$. The result is
\begin{eqnarray}
\dot{\pi}_m=(E_m^{-1}-1)g_{m}\pi_m+(E_m^1-1)m \pi_m,
\label{reducedmaster}
\end{eqnarray}
where $g_m =
\sum_{n=0}^{\infty}P_{n|m}(t)g_{m,n}$ is the production
rate of the plasmids averaged over the (yet unknown) conditional distribution
$P_{n|m}(t)$.

Further analytical progress is only possible in some limits.
Following Paulsson \textit{et al.} \cite{paulsson}, we assume that the $S$-dynamics is much faster than the $X$-dynamics.  At the level of the RRE, $S$ adjusts rapidly to the current value of $X$. Then $S(t)\simeq \eta X(t)$ holds, while $X(t)$ and $S(t)$ flow relatively
slowly towards the fixed point $(\bar{X},\bar{S})$ according
to the reduced equation $\dot{X}\simeq X \,\Psi(\eta X/A)-X$.
We will perform further calculations in two particular examples: exponential and hyperbolic inhibition.

\begin{figure}
\includegraphics[width=4.5cm,clip=]{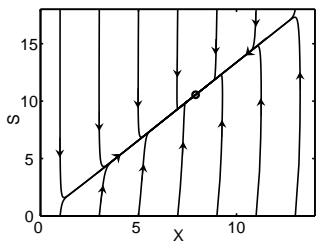}
\caption{The phase plane $X,S$ of the reaction rate equations
(\ref{meanfield}) for $\Psi(S/A)=k\exp(-S/A)$, $\beta\!=\!150$,
$\alpha\!=\!200$, $A=4$, and $k=14$. The circle denotes the fixed point
$(\bar{X},\bar{S})$.} \label{phaseplane}
\end{figure}

\textit{Exponential inhibition.} Here $\Psi(S/A)=k \exp(-S/A)$,
$k>1$ (we assume that $k$ is not too close to $1$), and $(\bar{X},\bar{S})=(A \ln k/\eta,\, A\ln k)$. The time scale of the fast dynamics is $\sim 1/\beta$,
the time scale of the slow dynamics is $\sim 1/\ln k$ [see Fig.~\ref{phaseplane}], so the time scale separation occurs when $\ln k \ll \beta$.

At the level of the CME we can perform adiabatic elimination of the
fast dynamics in the variable $P_{n|m}(t)$ by assuming that the
$S$-population rapidly adjusts to a Poisson distribution
$P^{(P)}_{n|m}=e^{-\eta m}(\eta m)^n/n!$ about the current value of
the mean, $\eta m(t)$. Now the effective stochastic rate
$g_m$ in Eq. (\ref{reducedmaster}) can be easily calculated
\cite{paulsson}:
\begin{equation}\label{effectiverate}
g_m=\sum_{n=0}^{\infty}P^{(P)}_{n|m}g_{m,n}=k m e^{-r\eta m},\;r=1-e^{-1/A}\,.
\end{equation}
This procedure reduces the two-species problem to an effective one-species problem:
a single-step birth-death process with the
birth rate $g_m$ and death rate $\mu_m=m$.  
In the most interesting case of $\langle m \rangle \gg 1$,
there are two widely different time scales in this process. The
first, short time scale is the relaxation time to the metastable state.
The second, exponentially long, time scale is the life time of the
metastable state, or the extinction time, see below. At intermediate
times one observes a \textit{quasistationary}
distribution (QSD) $q_m(t)=\pi_m/[1-\pi_0(t)]$: the PDF of having,
at time $t$, $m$ plasmids conditioned on their non-extinction, see
\textit{e.g.} \cite{nasell}. When $g_1/\mu_1\equiv k e^{-r\eta}\gg 1$,
that is  $\ln k \gg r\eta$, the probability flux to the zero
state $m=0$ is negligible, and $q_m(t)$ can be approximated
by putting, in Eq.~(\ref{reducedmaster}),
$\dot{\pi}_m(t)=0$ for all $m$ and assuming $\mu_1=0$ \cite{quasi}.
In this way one obtains a recursion relation for $q_m$
\cite{oppenheim,gardiner} which yields the QSD:
\begin{equation}\label{probdistexp}
\frac{q_m}{q_1}\simeq\frac{g_1 g_2\cdots g_{m-1}}{\mu_2\mu_3\cdots\mu_m}=\frac{e^{-(1/2)\,r\eta
m(m-1)}k^{m-1}}{m}\,,
\end{equation}
while $q_1$ can be found from the normalization
$\sum_{m=1}^{\infty}q_{m}=1$. Assuming $r \eta \ll 1$, we can
replace the normalization sum by an integral \cite{sumint} and, by the saddle point method, obtain
\begin{equation}\label{q1}
q_1^{-1} = \tau \simeq \frac{\sqrt{2\pi r\eta}}{\sqrt{k}\ln
k}\;e^{\frac{(\ln k)^2}{2r\eta}}\,.
\end{equation}
Now, $q_1^{-1}$ is nothing but the MET  $\tau$
\cite{nasell}, and Eq.~(\ref{q1}) yields an accurate
approximation for it. The same result follows from an
exact expression for the MET \cite{met}.

Let us calculate for comparison the MET for the
``semi-deterministic" (SD) case: when $S=\eta X$ is a prescribed
deterministic quantity. The SD rate $g_m^{sd}=k m e^{-\eta m/A}$ is
obtained by putting $n=\eta m$. A similar calculation, for $\eta/A
\ll 1$, yields
\begin{equation}\label{extincdet}
\tau^{sd}\simeq \frac{\sqrt{2\pi\eta}}{\sqrt{kA}\ln k}e^{\frac{A (\ln
k)^2}{2\eta}}\,.
\end{equation}
How do the fully stochastic (\ref{q1}) and SD (\ref{extincdet}) results
for the METs compare? Consider their ratio
\begin{equation}\label{extratio}
R\equiv\frac{\tau}{\tau^{sd}}=\sqrt{rA}\exp\left[\frac{(\ln
k)^2}{2r\eta}(1-rA)\right]\,.
\end{equation}
The strongest effect is observed for $(\ln k)^2 \gg \eta$. In this
case, and for $A \gg \eta$, we obtain $R\gg 1$: the discrete noise of the
$S$-molecules greatly (exponentially) delays the plasmid extinction. Note that in this parameter regime $R$ is a
monotone decreasing function of $A$. However, even for $A \to
\infty$ the effect is strong, as  $R \to e^{(\ln k)^2/(4\eta)}\gg
1$.

Using Eq.~(\ref{q1}) for $\tau$, we can determine the extinction
probability $\pi_0(t)$: the probability that extinction occurs until
time $t$, see \textit{e.g.} Ref. \cite{AM}. Also, by conservation of probability, we can restore the
exponentially slow time-dependence of the PDF of the metastable
state, $\pi_{m>0}(t)\simeq q_m\,\exp(-t/\tau)$ [where $\tau$ is
given by Eq.~(\ref{q1})]. We obtain
\begin{eqnarray}
\pi_0(t)&\!\simeq\!&1-e^{-t/\tau}\,,\nonumber\\
\hspace{-5mm}
\pi_{m>0}(t)&\!\simeq\!&\frac{k^{m-1/2}\ln k
}{m\sqrt{2 \pi r\eta}}e^{-\frac{(\ln k)^2}{2r\eta}-\frac{r\eta
m(m-1)}{2}-\frac{t}{\tau}} . \label{normprobdistexp}
\end{eqnarray}
Using the PDF (\ref{normprobdistexp}), we can calculate the (slowly
decaying in time) average number of the plasmids:
\begin{equation}\label{avg}
\langle
m(t)\rangle=\sum_{m=0}^{\infty}m\pi_m(t)\simeq\left(\frac{\ln
k}{r\eta}+\frac{1}{2}\right)e^{-t/\tau}\,,
\end{equation}
where we have again assumed $r \eta \ll 1$. For $A\lesssim 1$, $\langle
m(t)\rangle$ strongly deviates from the RRE prediction $\bar{X}=A
\ln k/\eta$, even at $t\ll \tau$, as was previously observed
numerically \cite{paulsson}. Note that non-gaussianity of the PDF
(\ref{normprobdistexp}) appears only in the pre-exponent.

\begin{figure}
\includegraphics[width=5.8cm,clip=] {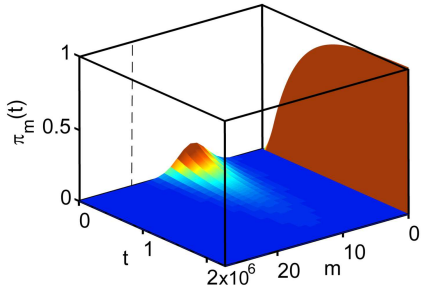}
\caption{(Color online) The PDF $\pi_m(t)$ found by solving
numerically a truncated CME (\ref{fullmaster}) for the exponential inhibition with $k=13$, $A=4$ and
$\alpha=\beta=400$. The dashed line shows the initial distribution:
a Kroenecker delta at $m=n=18$.} \label{prob3d}
\end{figure}

\begin{figure}
\includegraphics[width=5.3cm,clip=]{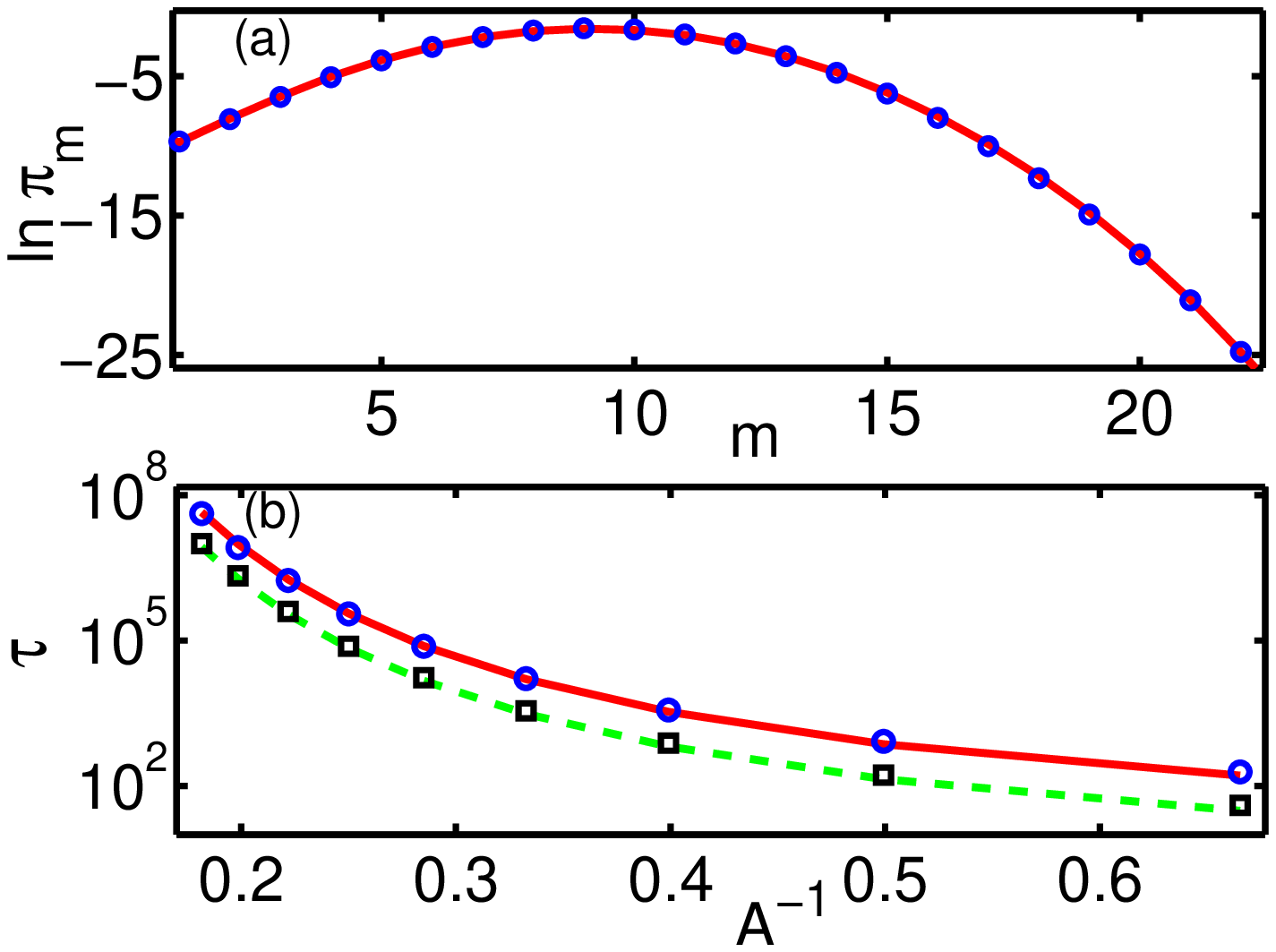}
\caption{(Color online) (a) The PDF
(\ref{normprobdistexp}) of the metastable state for the
exponential inhibition (solid line) and numerical solution of the
CME (\ref{fullmaster}) (circles) for $t\ll\tau$. The parameters are
$k=13$, $A=3$ and $\alpha=\beta=500$. (b) The MET versus $A^{-1}$,
the rest of parameters the same as in (a). Solid line: Eq.
(\ref{q1}), circles: numerical solutions for the fully stochastic
case, dashed line: Eq. (\ref{extincdet}), squares: numerical
solution for the SD case, see text for details. The ratio of the
fully stochastic and SD METs increases with the inhibition strength
$1/A$.} \label{combined}
\end{figure}

To test our analytical results, we solved numerically a truncated
CME (\ref{fullmaster}). The numerically found PDF of
the plasmids $\pi_m(t)$ 
exhibits a slow decay of the metastable state and
a simultaneous growth of the extinction probability in time, see Fig.~\ref{prob3d}.
Figure \ref{combined} compares our
analytical and numerical results for $\pi_m(t)$. In addition, we
compare there the analytical result (\ref{q1}) for the MET with the
numerical result $\tau^{num}=-t/\ln[1-\sum_{n=0}^{N}P_{0,n}(t)]$
(that approaches a constant after a transient), and also $\tau^{sd}$
from Eq. (\ref{extincdet}) with the result of a numerical solution
of Eq.~(\ref{reducedmaster}) with the SD rate $g_m$. Very good
agreement is observed for all quantities \cite{separation}.


\begin{figure}
\includegraphics[width=7.1cm,clip=]{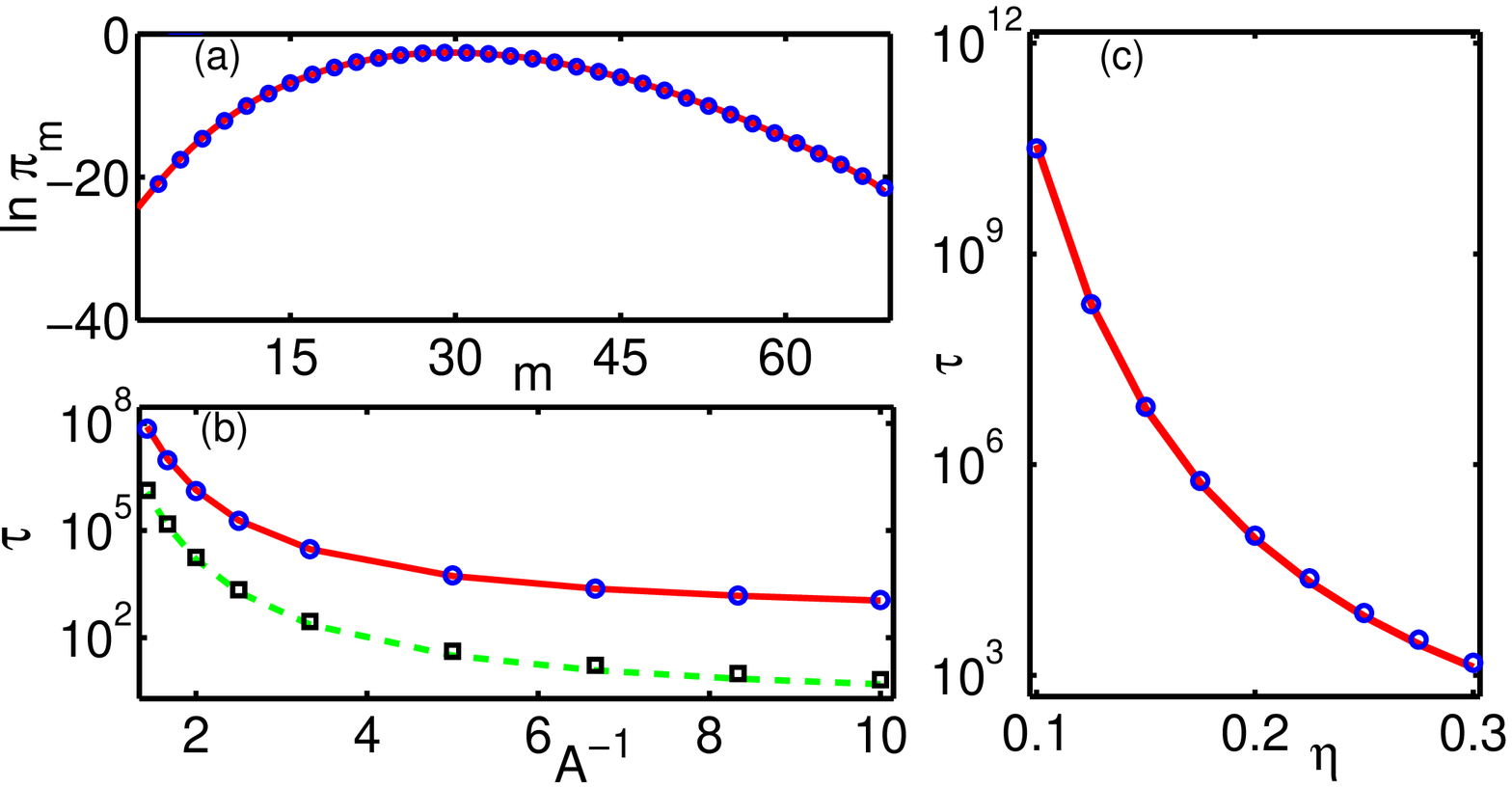}
\caption{(Color online) (a) The PDF
(\ref{fullprobdisthyp}) of the metastable state for the
hyperbolic inhibition (solid line) and numerical solution of the CME
(\ref{fullmaster}) (circles) for $t\ll\tau$. The parameters are
$k=15$, $A=1$, $\alpha=350$, and $\beta=700$. (b) The MET versus
$A^{-1}$, the rest of parameters the same as in (a). Solid line: Eq.
(\ref{fullprobdisthyp}), circles: numerical solutions for the fully
stochastic case, dashed line: Eq. (\ref{fullprobdisthyp}) with the
SD $g_m$, squares: numerical solution for the SD case, see text for
details. The ratio of the fully stochastic and SD METs grows
with the inhibition strength $1/A$. (c) Equation~(\ref{methyp}) for the MET  (solid line) and numerical solutions of the CME (\ref{fullmaster}) (circles) at  different $\eta$ for $k=10$, $A=10^{-3}$ and $\beta=2 \times 10^3$.}\label{exttime}
\end{figure}

\textit{Hyperbolic inhibition.} Our second example employs the
widely used hyperbolic, or Michaelis-Menten, inhibition model \cite{MM}. Here
$\Psi(S/A)=k/(1+S/A)$, $k>1$ (and not too close to
$1$), and $(\bar{X},\bar{S})=[(k-1)A/\eta,\, (k-1)A]$. The time
scale separation occurs at  $\beta>>1$. At the level of the CME
(\ref{fullmaster}) we again assume a rapid adjustment of the
$S$-species to the $m$-dependent Poisson distribution. The
effective stochastic rate $g_m$ in Eq.~(\ref{reducedmaster}) is
\begin{equation}
g_m=\sum_{n=0}^{\infty}P_{n|m}^{(P)}g_{m,n}=k m e^{-\eta
m}\,_1\!F_1(A,A+1,\eta m), \label{fullratehyp}
\end{equation}
where $_1\!F_1(a,b,z)$ is the Kummer confluent hypergeometric
function \cite{Abramowitz}. Using this effective rate, Paulsson and
Ehrenberg \cite{paulsson} calculated the QSD numerically. We have
found it analytically from the recursion relation \cite{oppenheim,gardiner}, by assuming
$g_1\gg \mu_1=1$. The result is
$q_m/q_1 \simeq (1/m!)\,\prod_{j=1}^{m-1}g_{j}$. Again, $q_1^{-1}=\tau$ can
be found by normalizing the QSD to unity. Therefore, the PDF of having $m$ plasmids at time $t$, and the
MET, are
\begin{equation}
\pi_{m>0}(t) \simeq \frac{e^{-t/\tau}}{\tau
m!}\,\prod_{j=1}^{m-1}g_{j}\,,\;\;\;\tau \simeq \sum_{m=1}^{\infty}\frac{1}{m!}\prod_{j=1}^{m-1}g_{j}\,.
\label{fullprobdisthyp}
\end{equation}
Comparisons between these predictions for the fully stochastic and
SD cases [with truncated sums in Eq.~(\ref{fullprobdisthyp})] and
numerical solutions of the truncated CMEs (\ref{fullmaster}) and
(\ref{reducedmaster}), respectively, are shown in
Fig.~\ref{exttime}a and b, and very good agreement is observed \cite{separation}.

The extreme case of very strong inhibition, $A\ll \ln k/k$, can be further simplified. It can
be checked \textit{a posteriori} that here, for
all $m$ that contribute to the normalization of $\pi_m$, and hence
to the MET, the effective rate (\ref{fullratehyp}) is well approximated
by the first
term: $g_m\simeq k m e^{-\eta m}$. This rate formally coincides with
that given by Eq.~(\ref{effectiverate}) for the exponential
inhibition, where one must put $r=1$. Therefore, the most
interesting case of strong inhibition $A\ll \ln k/k$ (when the
stabilizing effect of noise in the $S$-molecules on the plasmid
fluctuations and persistence is the largest) is also the simplest.
Furthermore,
the \textit{exponential} inhibition model
formally describes the strong \textit{hyperbolic}
inhibition limit. By additionally assuming that $A\ll \eta/k\ll \ln
k/k$, one can show after some algebra that the QSD and $\pi_m(t)$ from Eq.~(\ref{fullprobdisthyp})
reduce to Eqs.~(\ref{q1}) and
(\ref{normprobdistexp}), respectively (with $r=1$). The
slowly-decaying average of this PDF [Eq.~(\ref{avg}) with $r=1$]
again strongly deviates, already at $t\ll\tau$, from the RRE
prediction $\bar{X}=(k-1)A/\eta$. The corresponding MET
[Eq.~(\ref{q1}) with $r=1$] can again be compared with the SD MET,
obtained by using the SD rate $g_m^{sd}=k m (1+\eta m/A)^{-1}$ which
assumes $n=\eta m$. For $A\ll \eta/k$ one has $g_1^{sd}\ll 1$, so
$\tau_{sd}={\cal O}(1)$, as the decay dominates over the replication.
In contrast, the
asymptotic result for the stochastic MET, for $\eta \ll 1$,
\begin{equation}\label{methyp}
\tau \simeq \frac{\sqrt{2\pi \eta}}{\sqrt{k}\ln k}\;e^{\frac{(\ln
k)^2}{2\eta}}\,,
\end{equation}
is exponentially large. Therefore,  the noise in the number of the $S$-molecules
again causes,  at $A\ll \eta/k$, exponential enhancement of the persistence of the plasmids. Equation~(\ref{methyp}) compares well with numerics,
see Fig.~\ref{exttime}c.

\textit{Discussion.} We have shown, in a simple CNC model, that
intrinsic discrete noise of the signal molecules
can greatly increase the average number of regulated molecules and
therefore enhance the persistence of the regulated component.
Although we assumed that $P_{n|m}$ is Poisson distributed, we expect these
findings to hold, for sufficiently strong inhibition, for other signal molecule kinetics as well.
What is the mechanism behind the noise
enhanced persistence? The autocatalytic production rate of the
plasmids is largest at $S=0$, therefore rare events of having a very
small number of $S$-molecules strongly dominate the effective
stochastic growth rate $g_m$, see \textit{e.g.}
Eq.~(\ref{effectiverate}). As a result, the average number of
plasmids in the metastable state greatly increases, and this
enhances the plasmid persistence. As the mode and the average
for the plasmid PDF $\pi_m$ \textit{coincide}, this mechanism of failure
of the RRE is different from that discussed previously
\cite{samoilov}.

The noise-enhanced persistence, that we predict here, should be observable in experiment,
\textit{in vitro} and \textit{in vivo}, due to recent advances in
single-molecule signal measurements. Finally, the effect is not
system-specific, and should appear in a host of other
birth-death-type systems where negative feedback is at work.

We thank Ari Meerson for a useful discussion. Our work was supported
by the Israel Science Foundation.


\begin{thebibliography}{99}
\bibitem{quasi} When substituting $q_m(t)$ in Eq.~(\ref{reducedmaster}), one
obtains an equation which differs from Eq.~(\ref{reducedmaster}) by the presence of an additional term $q_1 q_m$ on the right side \cite{nasell}. For $g_1\gg \mu_1$, however, this term can be neglected.
\bibitem{sumint} The sum $\sum_m a_m$ can be replaced by an
integral if  $\delta_m=\left|1-a_{m+1}/a_{m}\right| \ll 1$ in the
region of a significant contribution to the sum. By virtue of the
saddle point method, this region is $m_*-\sigma\lesssim m \lesssim
m_*+\sigma$, where $m_*=\ln k/(r \eta)+1/2$ and
$\sigma=(r\eta)^{-1/2}$ are the mode and the standard deviation of
the QSD, respectively. Evaluating $\delta_m$, \textit{e.g.} at
$m=m_*+\sigma$, we obtain $\delta_m
\lesssim\left|1-e^{-\sqrt{r\eta}}\right|\ll 1$ which demands
$r\eta\ll 1$.
\bibitem{met} For the single-step birth-death process
(\ref{reducedmaster}), the MET, when starting from a state $m_0$, is
given by \cite{gardiner}: \begin{equation}\label{met} \tau(m_0)=
\sum_{i=0}^{m_0}\prod_{j=1}^{i}\frac{\mu_{j}}{g_j}
\sum_{l=i}^{\infty}\left(g_l\prod_{s=1}^{l}\frac{\mu_{s}}{g_s}\right)^{-1}\,.
\end{equation}
For the exponential inhibition the products can be easily
calculated: $\prod_{i=1}^{m} (\mu_{i}/g_i)=k^{-m}e^{r\eta
m(m-1)/2}.$ Then, for  $r\eta\ll 1$, both sums in (\ref{met}) can be
replaced by integrals. For $m_0\gg 1$ the inner and outer integrals
receive their main contributions from the vicinity of $l=\ln
k/(r\eta)$ and $i=0$, respectively. Therefore, we can use the saddle
point method for the inner integral and a Taylor expansion for the
outer one. The final result coincides with Eq. (\ref{q1}).
\bibitem{separation} Note that the CME may require a different criterion
of time scale separation between the $S$ and $X$
dynamics, compared to the RRE. To get an accurate \textit{a posteriori}  criterion, one can use the RRE with the effective stochastic
growth rate $g_m$. For the exponential inhibition this yields $\beta\gg \ln k$, as in the deterministic case.  For the hyperbolic inhibition one obtains $\beta\gg 1$ (as in the deterministic case) for $A\gtrsim \ln k/k$, and
$\beta\gg \ln k$ (as for the
exponential inhibition) for $A\ll\ln k/k$.
\end{thebibliography}
\end{document}